\documentclass[twocolumn,preprintnumbers,amsmath,amssymb]{revtex4}

\usepackage{graphicx}
\usepackage{dcolumn}
\usepackage{bm}

\begin{document}


\title{Characterizing and modeling citation dynamics}

\author{Young-Ho Eom}
\author{Santo Fortunato}%
\affiliation{Complex Networks and Systems Lagrange Laboratory,
Institute for Scientific Interchange (ISI), Viale S. Severo 65, 10133,
Turin, Italy}

\date{\today}

\begin{abstract}
Citation distributions are crucial for the analysis and modeling of
the activity of scientists.
We investigated bibliometric data of papers published in journals of
the American Physical Society, searching for the type of
function which best describes the observed citation distributions.
We used the goodness of fit with Kolmogorov-Smirnov statistics for
three classes of functions:
log-normal, simple power law and shifted power law. The shifted power
law turns out to be the most reliable hypothesis for all citation
networks we derived, which correspond to different time spans. We
find that citation dynamics is characterized by bursts, usually
occurring within a few years since publication of a paper, and the burst size spans
several orders of magnitude.
We also investigated the microscopic mechanisms for the evolution of
citation networks, by proposing a linear preferential attachment with time dependent initial attractiveness.
The model successfully reproduces the empirical citation distributions
and accounts for the presence of citation bursts as well.
\end{abstract}

\maketitle

\section{Introduction}

Citation networks are compact representations of the relationships
between research products, both in the sciences and the humanities~\cite{garfield55,garfield79}.
As such they are a valuable tool to uncover the dynamics of scientific
productivity and have been studied for a long time, since the seminal
paper by De Solla Price~\cite{price65}. In the last years, in
particular, due to the increasing availability of large bibliographic
data and computational resources, it is possible to build large networks and analyze them to an
unprecedented level of accuracy.

In a citation network, each vertex represents a paper and there is a
directed edge from paper $A$ to paper $B$ if $A$ includes $B$ in its
list of references. Citation networks are then directed, by
construction, and acyclic, as papers can only point to older papers,
so directed loops cannot be obtained.  A large part of the literature
on citation networks has focused on the characterization of the
probability distribution of the number of citations received by a
paper, and on the design of simple microscopic models able to
reproduce the distribution.
The number of citations
of a paper is the number of incoming edges (indegree) $k^{in}$ of the
vertex representing the paper in the citation network. So the
probability distribution of citations is just the indegree distribution $P\left(k^{in}\right)$.
There is no doubt that citation
distributions are broad, as there are papers with many citations
together with many poorly cited (including many uncited) papers.
However, as of today, the functional shape of citation distributions
is still elusive. This is because the question is ill-defined. In
fact, one may formulate it in a variety of different contexts, which
generally yield different answers. For instance, one may wish to
uncover the distribution from the global citation network including
all papers published in all journals at all times. Otherwise, one may
wish to specialize the query to specific disciplines or years.
The role of the discipline considered is important and is liable
to affect the final result. For instance, it is well known that papers
in Biology are, on average, much more cited than papers in
Mathematics. One may argue that this evidence may still be consistent
with having similar
functional distributions for the two disciplines, defined on ranges of
different sizes. Also, the role of time is important. It is unlikely
that citation distributions maintain the exact same shape regardless
of the specific time window considered. The dynamics of scientific
production has changed considerably in the last years. It is well
known, for instance, that the number of published papers per year has
been increasing exponentially until now~\cite{price75}. This, together
with the much quicker publication times of modern journals, has deeply
affected the dynamics of citation accumulation of papers. Moreover, if
the dataset at study includes papers published in different years,
older papers tend to have more citations than recent ones just because
they have been exposed for a longer time, not necessarily because they
are better works: the age of a paper is an important factor.

So, the question of which function best describes the citation
distributions is meaningless if one does not define precisely the set
of publications examined.
Redner~\cite{redner98} considered all papers published in Physical Review D up to $1997$,
along with all articles indexed by Thomson Scientific in the period $1981$-$1997$,
and found that the right tail
of the distribution, corresponding to highly cited papers, follows a
power law with exponent $\gamma=3$, in accord with the conclusions of Price~\cite{price65}.
Laherr\'ere and Sornette~\cite{laherrere98}
studied the top $1120$ most cited physicists
during the period $1981$-$1997$, whose citation distribution is more compatible with
a stretched exponential $P\left(k^{in}\right) \sim
\exp{  \left[ - \left(k^{in} \right)^{\beta} \right] } $,
with $\beta \simeq 0.3$.
Tsallis and de Albuquerque~\cite{tsallis00} analyzed
the same datasets used by Redner with an additional
one including all papers published up to $1999$ in Physical Review E, and
found that the Tsallis distribution
$P\left(k^{in}\right) = P(0) / \left[ 1+\left(\beta -1 \right) \, \lambda \, k^{in} \right]^{\beta/(\beta-1)}$, with $\lambda \simeq 0.1$ and $\beta \simeq 1.5$,
consistently fits the whole distribution of citations (not just the tail).
More recently Redner performed an analysis
over all papers published in the $110$ years long history
of journals of the American Physical Society (APS)~\cite{redner05},
concluding that the log-normal distribution
\begin{equation}
P\left(k^{in}\right) = \frac{1}{k^{in} \, \sqrt{2 \pi \sigma^2}} \; \exp{ \left\{ -\left[ \ln{\left(k^{in}\right)} - \mu \right]^2/\left(2 \sigma^2\right) \right\} } \;\;
\label{eq:lognormal}
\end{equation}
is more adequate than a power law.
In other studies distributions of citations have been fitted
with various functional forms:
power-law~\cite{seglen92, vazquez01, lehmann03, bommarito10, perc10, rodrigueznavarro11},
log-normal~\cite{stringer08, radicchi08, bommarito10},
Tsallis distribution~\cite{wallace09, anastasiadis10}, modified
Bessel function~\cite{vanraan01, vanraan01b} or more complicated
distributions~\cite{kryssanov07}.

In this paper we want to examine citation networks more in
depth. We considered
networks including all papers and their mutual citations within several
time windows.
We have performed a detailed analysis of the shape of the
distributions, by computing the goodness of fits with
Kolmogorov-Smirnov statistics of three model functions: simple power
law, shifted power law and log-normal. Moreover, we
have also examined dynamic aspects of the process of citation
accumulation, revealing the existence of ``bursts'', i.e. of rapid
accretions of the number of citations received by papers. Citation
bursts are not compatible with standard models of citation accumulation
based on preferential attachment~\cite{barabasi99}, in which the
accumulation is smooth and papers may attract many cites long after publication. Therefore, we
propose a model in which the citation attractiveness of a paper
depends both on the number of cites already collected by the paper and
on some intrinsic attractiveness that decays in time. The resulting
picture delivers both the citation distribution and the presence of bursts.

\section{Results}

\subsection{The distribution of cites}

For our analysis we use the citation database of the American Physical
Society (APS), described in Materials and Methods.
We get the best fit for the empirical citation distributions from
the goodness of fit test with Kolmogorov-Smirnov (KS)
statistics~\cite{clauset09}.
The KS statistic $D$ is the maximum
distance between the cumulative distribution function (CDF) of the
empirical data and the CDF of the fitted model:
\begin{equation}
D = \max_{k_{in} \geq k_{in}^{min}}|S(k_{in})-P(k_{in})|
\end{equation}
Here $S(k_{in})$ is the CDF of the empirical indegree $k_{in}$
and $P(k_{in})$ is the
CDF of the model that fits best the empirical data in the
region $k_{in} \geq k_{in}^{min}$.
By searching the parameter space, the best
hypothetical model is the one with the least value of $D$ from the empirical
data. To test the statistical significance of the hypothetical model,
we cannot use the values of the KS statistics directly though, as
the model has been derived from a best fit on the empirical data,
rather than being an independent hypothesis. So, following Ref.~\cite{clauset09}
we generate synthetic datasets from the model corresponding to
the best fit curve. For instance, if the best fit is the power law
$ax^{-b}$, the datasets are generated from this distribution. Each
synthetic dataset will give a value $D_{synth}$ for the KS statistics between
the dataset and the best fit curve. These $D_{synth}$-values are compared with
$D_{emp}$, i.e. the $D$-value between the original empirical data and the best fit
curve, in order to define a $p$-value.
The $p$-value is the fraction of $D_{synth}$-values
larger than $D_{emp}$. If $p$ is large (close to 1), the
model is a plausible fit to the empirical data; if it is close to $0$,
the hypothetical model is not a plausible fit.
We applied this goodness of fit test to three hypothetical model
distributions: log-normal, simple power law and shifted power law.
The log-normal distribution for the indegree $k_{in}$ is given by
\begin{equation}
P(k_{in}) \sim \frac{1}{k_{in}\sqrt{2\pi\sigma^2}}\exp\{-[\log(k_{in})-\mu]^{2}/(2{\sigma}^2)\},
\end{equation}
the simple power law distribution by
\begin{equation}
P(k_{in}) \sim {k_{in}}^{-\gamma},
\end{equation}
and the shifted power law by
\begin{equation}
P(k_{in}) \sim {(k_{in}+k_{0})}^{-\gamma} .
\end{equation}
We used $1000$ synthetic distributions to calculate the $p$-value for each empirical distribution.

\begin{figure*}[hbt!]
\begin{center}
\includegraphics[width=90mm, angle=-90]{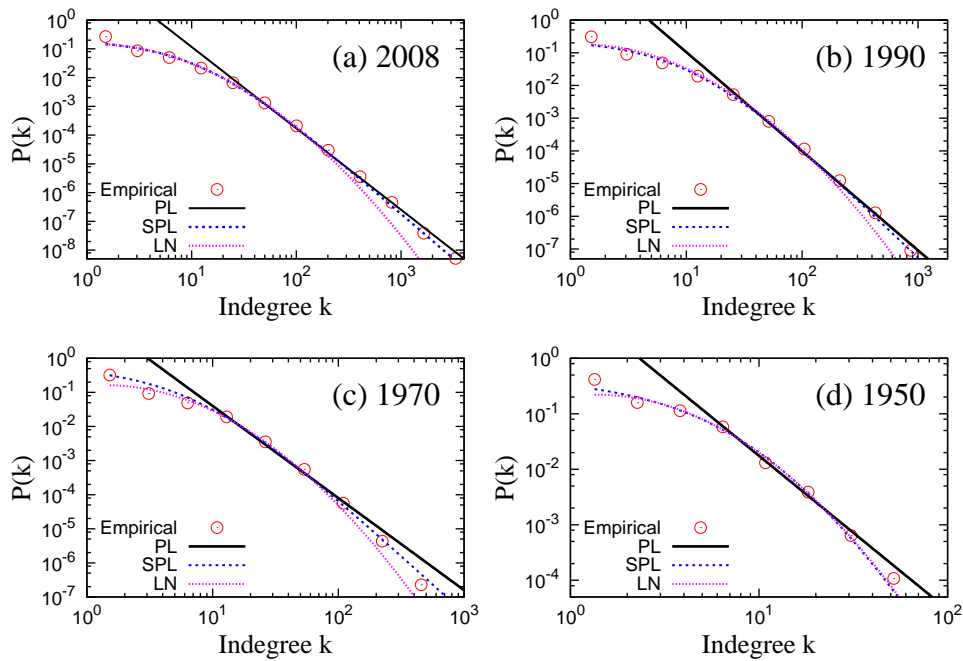}
\caption{Empirical citation distributions and best fit model distributions obtained through the goodness of fit with Komolgorov-Smirnov statistics. PL: Power law. SPL: Shifted power law. LN: Log-normal}
\label{fig:fitting}
\end{center}
\end{figure*}

\begin{table*}[bt]
\caption{\label{tab:fitting} Summary of the results of the goodness of fit test with Kolmogorov-Smirnov statistic on the empirical citation distributions for three test functions. log-normal (LN), simple power law (PL) and shifted power law (SPL).}
\begin{center}
\begin{tabular}{cccccccccccccc}
\hline \hline
distribution & 1950 & 1955 & 1960 & 1965 &1970 & 1975 & 1980 & 1985 & 1990 & 1995 & 2000 & 2005 & 2008 \\
\hline
LN  \\
p-value & 0.717 & 0.734 & 0.892 & 0.998 & 0.201 & 0.105 & 0.19 & 0.119 & 0.194 & 0.194 & 0.096 & 0.05 & 0.064 \\
$k_{min} $ & 2 & 3 & 7 & 14 & 2 & 2 & 2 & 3 & 2 & 2 & 2 & 2 & 2 \\
\hline
PL \\
p-value & 0.001 & 0.955 & 0.056 & 0.321 & 0.022 & 0.127 & 0.204 & 0.784 & 0.686 & 0.412 & 0.362 & 0.619 & 0.44 \\
$k_{min} $       & 6 & 16 &  9 & 19 & 12& 17 &20 &39 &46 &39 &43 &47 & 47   \\
\hline
SPL  \\
p-value & 0.832 & 0.777 & 0.49 & 1.00 & 0.943 & 0.958 & 0.49 & 0.728 & 0.909 & 1.00 & 0.797 & 0.989 & 0.99 \\
$k_{min} $              & 2 & 2 & 2 & 14 & 9 & 12 & 2 & 2 & 2 & 2 & 3 & 6 & 5 \\
\hline \hline
\end{tabular}
\end{center}
\end{table*}

Figs. 1a, 1b, 1c and 1d show some fits for datasets corresponding to several time
windows (see Materials and Methods). The detailed summary of the goodness of fit results is shown in
Table 1.
The simple power law gives high $p$-value only when one considers the right
tail of the distribution (usually $k_{in}>20$). The log-normal
distribution gives high $p$-value for early years (before 1970) but
after 1970 the p-value is smaller than 0.2. As shown in Figs.
1a and 1b, there is a clear discrepancy in the tail between the best fit
log-normal distribution and the empirical distribution. The
shifted power law distribution gives significant p-values (higher than
0.2) for all observation periods. The values of the exponent $\gamma$
of the shifted power law are decreasing in time.
The range of $\gamma$ goes from $5.6$ ($1950$) to $3.1$ ($2008$).

We conclude that the shifted power law is the best distribution to fit the data.

\subsection{The distribution of citation bursts}

We now turn our attention to citation ``bursts''. While there has been
a sizeable activity in the analysis of bursty behavior in human
dynamics~\cite{barabasi05,vazquez05,vazquez06}, we are not aware of
similar investigations for citation dynamics.
We compute the relative rate $\Delta k/k = [k(t+\delta
t)_{in}^{i}-k(t)_{in}^{i}]/k(t)_{in}^{i}]$, where $k(t)_{in}^{i}$ is the number
of citations of paper $i$ at time $t$.
The distributions of $\Delta k/k$
with $t=1949$, $1969$, $1989$, $2007$ and $\delta t=1$ year are shown
in Fig. 2a. They are visibly broad, spanning several orders of
magnitude. Similar heavy tails of burst size
distributions were observed in the dynamics of popularity in Wikipedia and
the Web~\cite{ratkiewicz10}.  It is notable that the largest bursts
take place in the first years after publication of a paper. This is
manifest in Fig. 2b, where we show distributions derived from the same dataset as in Fig. 2a,
but including only papers older than $5$ (squares) and
$10$ years (triangles): the tail disappears. In general, more
than $90\%$ of large bursts ($\Delta k/k > 3.0$) occur within the
first $4$ years since publication.

\begin{figure*}[hbt!]
\begin{center}
\includegraphics[width=55mm, angle=-90]{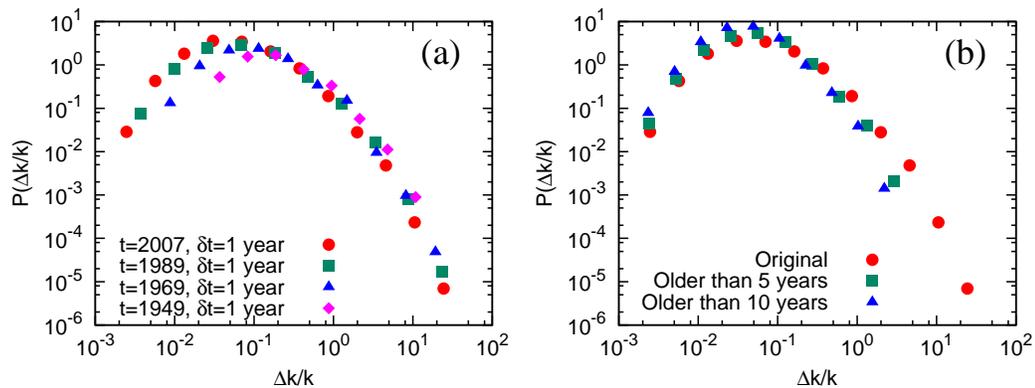}
\caption{(a) The four curves correspond to 1949, 1969, 1989 and 2007, the observation window is $\delta$t=1
year. (b) Here the reference year is 2007, but the burst statistics is limited to the papers published until 2003 (squares) and 1998 (triangles). For comparison, the full curve comprising all papers (circles, as in (a)) is also shown.}\label{fig:Burst}
\end{center}
\end{figure*}

\subsection{Preferential attachment and age-dependent attractiveness}

For many growing networks,
cumulative advantage~\cite{yule25,simon57}, or preferential
attachment~\cite{barabasi99}, has proven to be a reliable mechanism to
explain the fat-tailed distributions observed. In the context of
citation dynamics, it is reasonable to assume that, if a paper is very
cited, it will have an enhanced chance to receive citations in the future
with respect to poorly cited papers. This can be formulated by stating
that the
probability that a  paper gets cited
is proportional to the number of citations it already received. That was the original idea of
Price~\cite{price76} and led to the development of the first
dynamic mechanism for the generation of power law distributions in
citation networks. In later refinements of the model, one has
introduced an {\it attractiveness} for the vertices, indicating their
own appeal to attract edges, regardless of degree. In particular, one
has introduced the so-called {\it linear preferential
attachment}~\cite{krapivsky00,dorogovtsev00}, in which the probability
for a vertex to receive a new edge is proportional to the sum of the
attractiveness of the vertex and its degree.
In this Section we want to check whether this hypothesis holds for our
datasets. This issue has been addressed in other works on citation
analysis, like Refs.~\cite{wang08b,perc10}.

We investigated the dependence of the kernel function $\Pi(k_{in})$ on
indegree $k_{in}$~\cite{jeong03,eom08}. The kernel is the rate with
which a vertex $i$ with indegree $k_{in}^i$ acquires new incoming edges.
For linear preferential attachment the kernel is
\begin{equation}
\Pi(k_{in}^i) = \frac{k_{in}^i+A_i}{\sum_j [k_{in}^j+A_j]}.
\label{eqkern}
\end{equation}
In Eq.~\ref{eqkern} the constant $A_i$ indicates the attractiveness of
vertex $i$. Computing the kernel directly for each indegree class
(i.e. for all vertices with equal indegree $k_{in}$) is not ideal, as
the result may heavily fluctuate for large values of the indegree, due
to poor statistics. So, following Refs.~\cite{jeong03,eom08}, we
consider the cumulative kernel
$\Pi_>(k_{in}^i)=\sum_{k^\prime \leq k_{in}}\Pi(k^\prime)$, which, for the
ansatz of Eq.~\ref{eqkern}, should have the following functional
dependence on $k_{in}$
\begin{equation}
\Pi_>(k_{in}) \sim k_{in}^2+\langle A\rangle k_{in}.
\label{eqkerncum}
\end{equation}
In Eq.~\ref{eqkerncum} $\langle A\rangle$ is the average
attractiveness of the vertices. In order to estimate $\Pi_>(k_{in})$,
we need to compute the probability that vertices with equal indegree
have gotten edges over a given time window, and sum the results over
all indegree values from the smallest one to a given $k_{in}$. The
time window has to be small enough in order to preserve the structure
of the network but not too small in order to have enough citation
statistics. In Fig. 3 we show the cumulative kernel function $\Pi_>(k_{in})$ as a function
of indegree for a time window from $2007$ to $2008$. The profile of
the curve (empty circles) is compatible with linear preferential attachment with an average
attractiveness $\langle A\rangle = 7.0$ over a large range, although
the final part of the tail is missed. Still, the slope of the tail, apart from the final
plateau, is close to $2$, like in Eq.~\ref{eqkerncum}. Our result is
consistent with that of Jeong et al.~\cite{jeong03}, who considered a
citation network of papers published in Physical Review Letters in
1988, which are part of our dataset as well. We have repeated this
analysis for several datasets, from $1950$ until $2008$, by keeping a time
window of one year in each case. The resulting values of $\langle
A\rangle$ are reported in Table 2, along with the number of vertices
and mean degree of the networks.
The average value of the attractiveness across all datasets is
$7.1$. This value is much bigger than the average
indegree in the early ages of the network like, for example, from $1950$ to
$1960$. Hence, in the tradeoff between indegree and attractiveness of
Eq.~\ref{eqkern}, the latter is quite important for old papers.
In general, for low indegrees, attractiveness dominates over
preferential attachment. As we see in Fig. 3, in fact, for low
indegrees there is no power law dependence of the kernel on indegree.

\begin{figure*}[hbt!]
\begin{center}
\includegraphics[width=75mm, angle=-90]{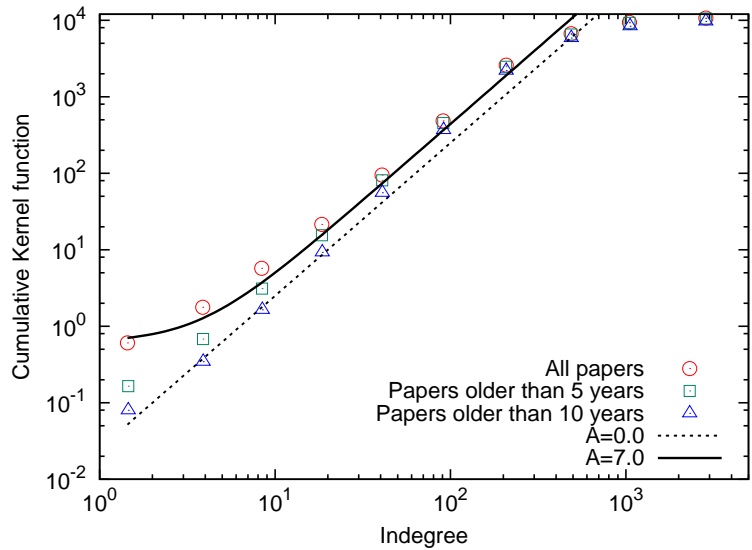}
\caption{ Cumulative kernel function of citation network from 2007 to 2008. The continuous line is $Ck_{int}(k_{int}+A)$ with A=7.0 and C is constant. The dashed line corresponds to the case without attractiveness ( $A=0.0$ ).}\label{fig:PA}
\end{center}
\end{figure*}

\begin{table*}[t]
\caption{\label{tab:parameter} Statistics of the empirical citation networks: N is the number of vertices in the network; $<k>$ is the average indegree of the network; $<A>$ is the average attractiveness, determined from the tests of linear preferential attachment discussed in the text.}
\begin{center}
\begin{tabular}{cccccccccccccc}
\hline \hline
 & 1950 & 1955 & 1960 & 1965 &1970 & 1975 & 1980 & 1985 & 1990 & 1995 & 2000 & 2005 & 2008 \\
\hline
$N$ & 15880 & 23350 & 30996 & 42074 & 62382 & 85590 & 108794 & 138206 & 180708 & 238142 & 305570 & 386569 & 441595 \\
$<k>$ & 2.2 & 3.1 & 3.7 & 4.3 & 5.1 & 5.6 & 6.0 & 6.2 & 6.5 & 7.0 & 7.7 & 8.5 & 9.0  \\
$<A>$ & 4.2 & 5.3 & 6.2 & 5.4 & 7.2 & 7.9 & 7.8 & 9.0 & 7.4 & 7.3 & 6.8 & 6.4 & 7.0\\
\hline \hline
\end{tabular}
\end{center}
\end{table*}

Finally we investigated the time dependence of the kernel. As shown in Fig. 3,
when we limit the analysis to papers older than 5 years (squares) or 10
years (triangles), the kernel has a pure quadratic dependence on indegree
in the initial part, without linear terms, so the
attractiveness does not affect the citation dynamics. This means that
the attractiveness has a significant influence on the evolution of
the citation network only within the first few years after publication
of the papers. The presence of vertex attractiveness had been
considered by Jeong et al. as well~\cite{jeong03}.

\subsection{The model}

We would like to design a microscopic model that reflects the observed
properties of our citation networks.
Preferential attachment does not account
for the fact that the probability to receive
citations may depend on time. In the Price model, for instance,
papers keep collecting citations independently of their age,
while it is empirically
observed~\cite{hajra04,hajra05,wang08b} that the probability for an article
to get cited decreases as the age of the same article increases. In
addition, we have seen that citation bursts typically occur in the
early life of a paper.
Some sophisticated growing network
models include the aging of vertices as well~\cite{dorogovtsev00b, dorogovtsev01b, zhu03, hajra05, wang08b}.
We propose a mechanism based on linear preferential attachment,
where papers have individual values of the attractiveness, and the
latter decays in time.

\begin{figure*}[tb!]
\begin{center}
\includegraphics[width=75mm,angle=-90]{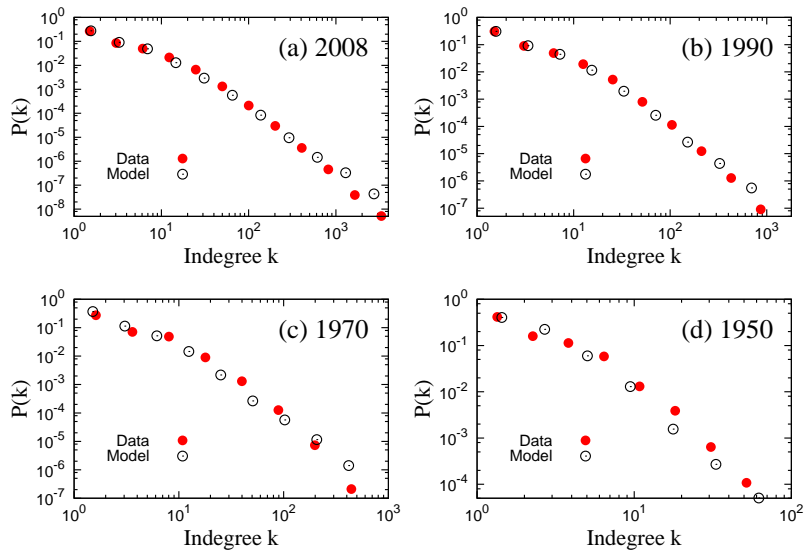}
\caption{Comparison of the citation distributions from the empirical data and our model. For all cases, we used $\alpha=2.5$ and $\tau=1$year. (a) For 2008, $N$=4415905, $<k>=9.0$ (b) For 1990, $N$=180708, $<k>=6.5$ (c) For 1970, $N$=62382, $<k>$=5.6 (d) For 1950, $N$=1950, $<k>$=3.1. Here $N$ is the number of papers/vertices and $<k>$ is the average number of citations/indegree. }
\label{fig:model}
\end{center}
\end{figure*}

The model works as follows. At each time step $t$, a new vertex
joins the network (i.e., a new paper is published). The new
vertex/paper has $m$ references to existing vertices/papers. The
probability $\Pi (i\rightarrow j, t)$ that the new vertex $i$ points
to a target vertex $j$ with indegree $k_{in}^j$ reads
\begin{equation}
\Pi (i\rightarrow j, t) \sim [k_{in}^j+A_j(t)],
\label{eq:kernel}
\end{equation}
where $A_j(t)$ is the attractiveness of $j$ at time $t$. If $A_j(t)$ were
constant and equal for all vertices we would recover the standard
linear preferential attachment~\cite{dorogovtsev00,krapivsky00}. We
instead assume that it decays exponentially in time
\begin{equation}
A(t) = A_{0} \exp [-(t-t_0)/\tau].
\label{eqattr}
\end{equation}
In Eq.~\ref{eqattr} $A_0$ is the initial attractiveness of the vertex,
and $t_0$ is the time in which the vertex first appears in the
network; $\tau$ is the time scale of the decay, after which the
attractiveness lowers considerably and loses importance for citation dynamics.
Since citation bursts occur in the
initial phase of a paper's life (Fig. 2b), when vertex attractiveness
is most relevant, we expect that the values of the initial attractiveness are
heterogeneously distributed, to account for the broad distribution of
burst sizes (Fig. 2a). We assume the power law distribution
\begin{equation}
P(A_{0}) \sim A_{0}^{-\alpha}.
\label{eq:adis}
\end{equation}
We performed numerical simulations of the model with parameters
obtained from the empirical data. We use $\alpha=2.5$, $\tau=1$ year and
$ A_{min} \leq A_0 < 0.002N(t)$ with $N(t)$ is the number of papers at
time $t$. The upper bound represents the largest average indegree of
our citation networks, expressed in terms of the number of vertices.
The value of $A_{min}$ depends on the obtained value of the attractiveness
from empirical data. We set $A_{min}=25.0$ for most years, for
1950 we set $A_{min}=14.5$, because $\langle A\rangle$ is smaller
than $7.1$. The result is however not very sensitive to the minimum and maximum
value of $A_0$. Figs. 4a, 4b, 4d and 4d show
the citation distributions of empirical data versus the model
prediction. The model can reproduce the empirical distributions very well
at different phases in the evolution of the APS citation network, from
the remote $1950$ (panel d) until the very recent $2008$ (panel a).

\begin{figure*}[htb!]
\begin{center}
\includegraphics[width=45mm, angle=-90]{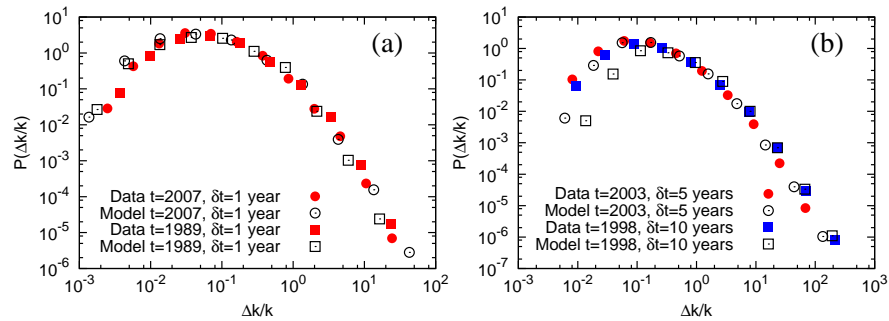}

\caption{Comparison of the distributions of citation burst size from the empirical data and the model. The exponent $\alpha$ of the distribution of initial attractiveness is 2.5, as in Fig. 4. (a) The reference years are 1989 (squares) and 2007 (circles), the observation window for the bursts is $\delta t=1$ year in both cases. (b) Here the reference years are 1998 (squares) and 2003 (circles) and the observation windows for the bursts are of
10 and 5 years, respectively.}
\label{fig:modelburst}
\end{center}
\end{figure*}

The distributions of citation burst magnitude $\Delta k/k$ for the data and the
model are shown in Fig. 5a. For a better comparison between data and model
we ``evolve'' the network according to the model by starting from the structure of the empirical citation
network at the beginning of the time window for the detection of the
bursts. We stop the evolution after the observation time $\delta t$
elapses. In Fig. 5a we consider $1989$ and $2007$, with a time window
of $1$ year for the burst detection. The model successfully reproduces
the empirical distributions of burst size. In Fig. 5b we consider much longer observation
periods for the bursts, of $5$ and $10$ years. Still, the model gives
an accurate description of the tail of the empirical curve in both
cases.

\section{Discussion}

We investigated citation dynamics for networks of papers published on journals of
the American Physical Society. Kolmogorov-Smirnov statistics along
with goodness of fit tests make us conclude that the best ansatz for
the distribution of citations (from old times up to any given year) is a shifted
power law. The latter beats both simple power laws, which are acceptable
only on the right tails of the distributions, and log-normals, which
are better than simple power laws on the left part of the curve, but
are not accurate in the description of the right tails. We have also
studied dynamic properties of citation flows, and found that the early
life of papers is characterized by citation bursts, like already found
for popularity dynamics in Wikipedia and the Web.

The existence of bursts is not compatible with traditional models
based on preferential attachment, which are capable to account for the
skewed citation distributions observed, but in which citation
accumulation is smooth.
Therefore we have introduced a variant of linear preferential
attachment, with two new features: 1) the attractiveness decays
exponentially in time, so it plays a role only in the early life of
papers, after which it is dominated by the number of citations accumulated; 2) the attractiveness is not the same for all vertices but it
follows a heterogeneous (power-law) distribution. We have found that
this simple model is accurate in the description of the distributions
of citations and burst sizes, across very different scientific
ages. Moreover, the model is fairly robust with respect to the choice
of the observation window for the bursts.

\section{Materials and Methods}

Our citation database includes
all papers published in journals of the American Physical Society (APS)
from 1893 to 2008, except papers published in Reviews of Modern
Physics. There are 3\,992\,736 citations among 414\,977 papers at the end of
2008. The journals we considered are Physical Review (PR), Physical
Review Letters (PRL), Physical Review A (PRA), Physical Review B
(PRB), Physical Review C (PRC), Physical Review D (PRD), Physical
Review E (PRE), Physical Review - Series I (PRI), Physical Review
Special Topics - Accelerators and Beams (PRSTAB), and Physical Review
Special Topics - Physics Education Research (PRSTPER). From these data, we constructed
time-aggregated citation networks from 1950 to a year $x$, with
$x=1951, 1952, ...., 2007, 2008$.

\section{Acknowledgments}

We thank the American Physical Society for letting us use their
citation database.

\bibliographystyle{apsrev4-1}
\bibliography{science} 

\newpage
\clearpage

\end{document}